\documentclass[showpacs,aps,prb,twocolumn,amsmath,amssymb,superscriptaddress]{revtex4}
\usepackage{graphicx}
\usepackage{graphics}
\usepackage{dcolumn}
\usepackage{bm}
\usepackage{amssymb,amsmath}
\newcommand{\beq}{\begin{equation}}
\newcommand{\eeq}{\end{equation}}
\newcommand{\beqnar}{\begin{eqnarray}}
\newcommand{\eeqnar}{\end{eqnarray}}
\newcommand{\bfig}{\begin{figure}}
\newcommand{\efig}{\end{figure}}

\begin{document}
\title{Metallic phase of disordered graphene superlattices with long-range correlations}

\author{Hosein Cheraghchi}
\email{cheraghchi@dubs.ac.ir}
\affiliation{School of Physics, Damghan University of Basic
Sciences, 6715- 364, Damghan, Iran}
\author{Amir Hossein Irani}
\affiliation{School of Physics, Damghan University of Basic
Sciences, 6715- 364, Damghan, Iran}
\author{Sayyed Mahdi Fazeli}
\affiliation{Physics Department, the
University of Qom, Qom, Iran}
\author{Reza Asgari}
\email{asgari@ipm.ir}
\affiliation{School of Physics, Institute
for Research in Fundamental Sciences, IPM 19395-5531 Tehran,
Iran}

\date{\today}
\newbox\absbox

\begin{abstract}
Using the transfer matrix method, we study the conductance of the
chiral particles through a monolayer graphene superlattice with
long-range correlated disorder distributed on the potential of
the barriers. Even though the transmission of the particles through
graphene superlattice with white noise potentials is suppressed, the transmission is revived in a wide range
of angles when the potential heights are long-range correlated
with a power spectrum $S(k)\sim1/k^{\beta}$. As a result,
the conductance increases with increasing the correlation exponent values gives rise a
metallic phase. We obtain a phase transition diagram
in which a critical correlation exponent depends strongly on
disorder strength and slightly on the energy
of the incident particles. The phase transition, on the other hand,
appears in all ranges of the energy from propagating to evanescent mode
regimes.
\end{abstract}


 \maketitle
\section{Introduction}
The exploration of graphene, a monolayer of carbon atoms tightly
packed into a honeycomb lattice, has been recently attracted
special attentions for investigation of the fundamental physics
and also probable device applications such as nanoelectronic based
on planar graphene structures~\cite{novoselov, geim}. In graphene,
due to its unique band structure with the valence and conduction
bands touching at two inequivalent Dirac points electrons around
the Fermi level obey the massless relativistic Dirac equation,
which results in a linear energy dispersion
relation~\cite{Dirac}. Massless relativistic quasiparticles
arising from the cone spectrum lead to a number of unusual
electronic properties such anomalous integer~\cite{QHE} and
fractional~\cite{kim} quantum Hall effects, focusing of electron
by a rectangular potential barrier (Veselago
lensing)~\cite{veselago}, special Andreev
reflection~\cite{Andreev}, observation of the plasmaron
composite~\cite{plasmaron} and minimal
conductivity~\cite{minimal}.

\bfig
\includegraphics[width=9 cm]{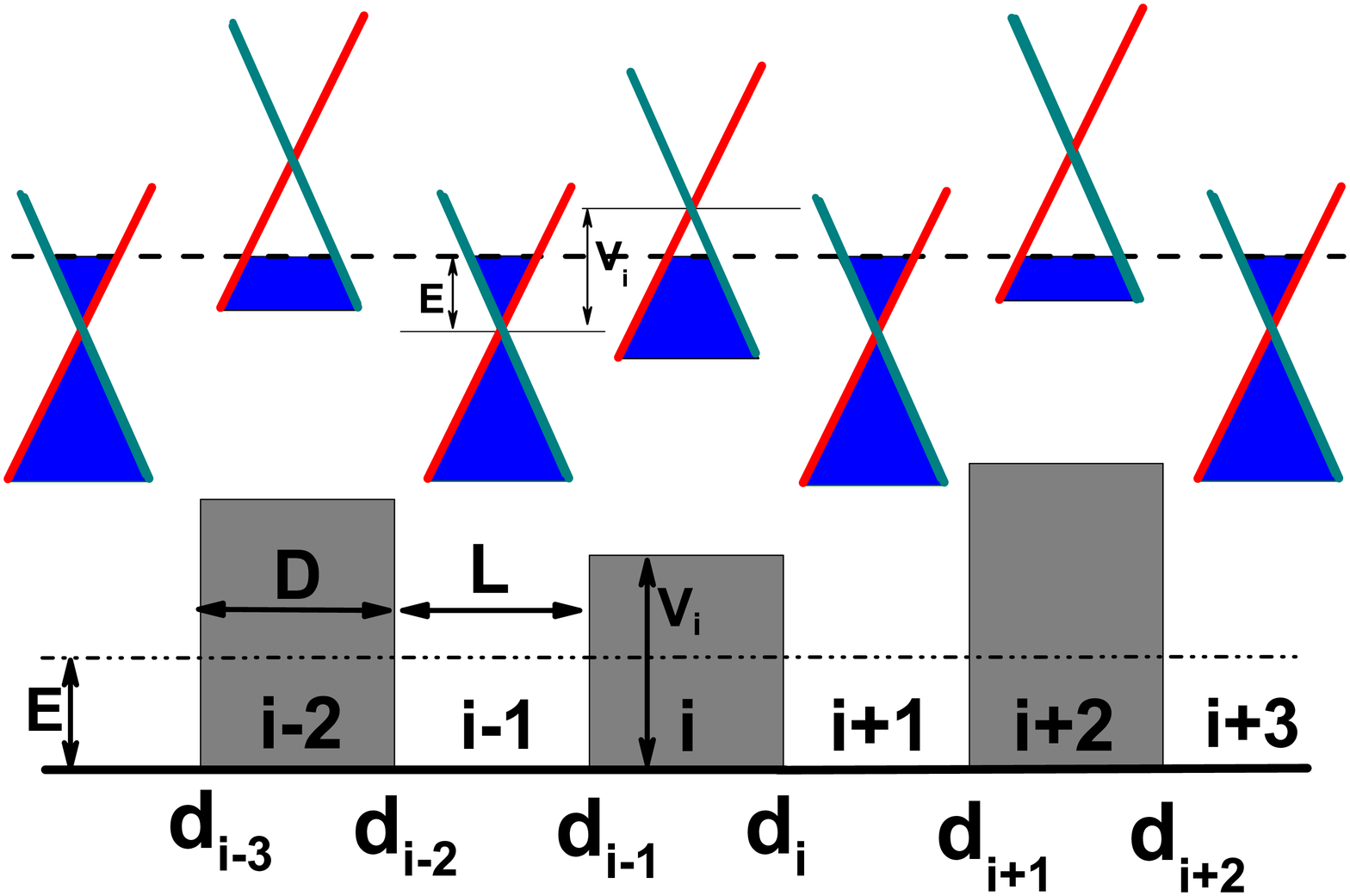}
\caption{Graphene superlattice with long-range correlated disorder
 on the potential barriers.}\label{superlattice} \efig

Interestingly, relativistic quantum quasiparticles incident normally to a high electrostatic
potential barrier in graphene can pass through it with perfect
transmission regardless of the height and width of the
barrier~\cite{klein}. This phenomenon which is referred as the Klein
tunneling is in contrast with the quantum massive carrier tunneling where the transmission
probability decays exponentially with increasing of the barrier
height and width. Recently, evidences for the Klein tunneling of
the Dirac fermions across p-n junction have been experimentally
observed when a gate-induced potential step is steep
enough~\cite{exp-klein1,exp-klein2}.

In graphene sheets the type of particle (electrons or holes)
and the density of the carriers can be controlled by tuning a gate
bias voltage~\cite{1,2,3}. Moreover, graphene superlattices may be fabricated
by adsorbing adatoms on the graphene surface by positioning and
aligning impurities with scanning tunneling microscopy~\cite{4}, or by
applying a local top gate voltage to graphene~\cite{5}. The transition
of hitting massless particles in a clean~\cite{zhang} or disordered~\cite{asgari}
graphene-based superlattice structure has been studied. It is
shown that the conductivity of the system depends on the
superlattice structural parameters.

The first study on electronic properties of monolayer and bilayer
graphene superlattices was performed by Bai and
Zhang~\cite{zhang}. They showed that the angularly averaged
conductivity can be controlled by changing the structure
parameters. It has been shown that massless Dirac fermions are
generated in one-dimensional external periodic potential close to
the original Dirac point~\cite{park,brey}. The Dirac points
depend on geometrical parameters for instance the potential of the
barriers/wells, the period of the potential and transverse wave
number~\cite{Barbier}. An evidence for such Dirac points is the
conductance resonances are appeared in the special potential
values~\cite{brey}. Moreover, the conductance of graphene
superlattice with uncorrelated disorder on the width of the
barriers was calculated in [~\onlinecite{asgari}]. It was shown
that the transmission of the quasiparticles with large angles
incidence to the potential barriers is suppressed by disorder
strength and the sample size too. Therefore, the results of the
finite-size scaling computations predicted a zero conductance for
all the graphene superlattices, except for some resonant barrier
thickness in which the conductance tends to a nonzero constant in
the thermodynamic limit~\cite{asgari}.

A number of numerical calculations of electron transport confirmed the absence of the
localization in the presence of the long-range random potential in
disordered graphene~\cite{nomura}.
The main quantity mostly studied numerically is
the conductance, $G$ of a finite-size graphene sample with a width,
$W$ much larger than the length, $L$. The setup allows us to
define the "conductivity" $\sigma = GL/W$ even for ballistic
samples with $L$ much shorter than the mean free path, $l$.

It is well known that the transmission of the quantum massive
carriers unexpectedly increases when special correlation is
applied on disorder~\cite{dunlap}. This is in contrast with the
Anderson localization in which all states are exponentially
localized in one dimensional uncorrelated disorder. Experimental
evidence for discrete the number of the extended states has been
observed in random-dimer semiconductor superlattices as a
short-range correlated disorder~\cite{experiment}. Long-range
correlated sequences of the potential barriers in semiconductor
superlattices, however, could result in a continuum of extended
states giving rise the mobility
edges~\cite{mouraprl,cheraghchi,esmailpour}.

In this paper, we study the conductance of massless Dirac fermions
through graphene superlattices with a long-range correlated disorder
on the potentials of the barriers. Transmission of the large angles
incident electrons to graphene superlattice which are suppressed
by uncorrelated disorder, are revived by applying correlation
between random potentials of the barriers. As a result, the conductance
increases with the correlation exponent. Consequently, an
insulator-to-metal transition emerges at a critical correlation
exponent which depends strongly on the disorder strength. One
should notice that such phase transition emerges for all ranges of
the energy. In addition, the dependence of the conductance to the
superlattice parameters is investigated for different correlation
strengths.

The paper is organized as follows: we present the transfer matrix
and Fourier Filtering method used to calculate transmission
through long-range correlated graphene superlattices in Section
II. Our results and discussions will be presented in Section IV
presenting the metal-to-insulator phase transition and investigating
the emergence of the phase transition along different energy
ranges. The last Section concludes our results.

\section{Model and the transfer Matrix}
\bfig
\includegraphics[width=9 cm]{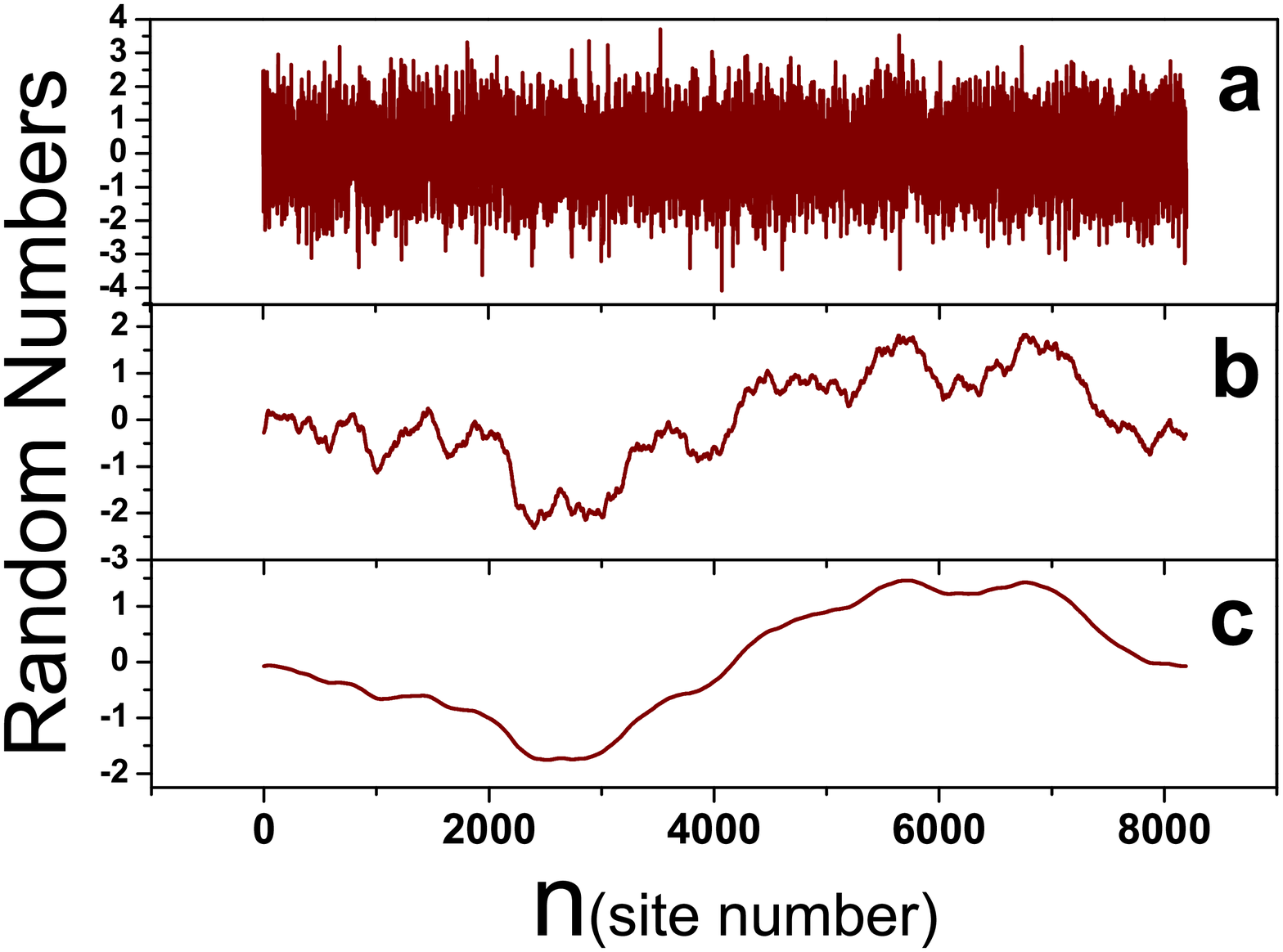}
\caption{Random distribution of the correlated sequences generated
by Fourier Filtering Method. a) uncorrelated case corresponding to
$\alpha=0.500$, b) $\alpha=1.766$, c)
$\alpha=1.993$.}\label{correlated-data} \efig

In the low-energy limit, charge carriers near the Dirac point in
the continuum model can be described by the following
non-interacting Hamiltonian

\beq H=-i \hbar v_{\rm F}
\overrightarrow{\sigma}.\overrightarrow{\nabla}+V(x)
  \label{dirac}
\eeq
where $v_{\rm F}=10^6$ m/s is the Fermi velocity and
$\overrightarrow{\sigma}=(\sigma_x,\sigma_y)$, are Pauli matrices.
We consider a lattice of the electrostatic potentials as barriers which are induced
by top gate voltages. Therefore, the potential of the barriers are
sorted as the following:

\beq V(x)= \left\{ \begin{array}{c} V_i \,\,\,\,\,\,\,\,\,\,\,
d_{2i-1}<x<d_{2i}, \,\,\,\,\,\,\ i=1,2,...
\\ \\ 0
\,\,\,\,\,\,\,\,\,\,\,\,\,\,\,\,\,\,\,\,\,\,\ {\rm
elsewhere}\end{array} \right. \label{potential}
 \eeq

A schematic representation of graphene superlattice is shown in
Fig.~\ref{superlattice}. We consider the number of regions
(wells and barriers) to be equal $N$ and then there is $(N-1)/2$
barriers. The width of the barriers and wells are considered to be
fixed. The height of the barriers fluctuates around its mean is
defined by $V_i= \langle V\rangle(1+\sigma~\varepsilon_i)$ where
$\sigma$ is the variance of the potentials of the barriers and
$\{\varepsilon_i\}$ is a long-range correlated random sequence of
data with the Gaussian distribution. The random sequences will
become normalized, accordingly the mean value $\varepsilon_i$ is
set to be zero and its variance has been fixed.

Before passing to calculate the conductance, we might generate a
random sequence with long-range correlation, $\{\varepsilon_i\}$
will be considered to describe the trace of a fractional Brownian
motion with a power spectrum $S(k)\sim1/k^{\beta}$ where $1<\beta<3$
and $\beta=2H+1=2\alpha-1$ ($1<\alpha<2$). Here $H$ is the Hurst
exponent. In the case of power-law decaying auto-correlations, the
correlation function decays with an exponent $\gamma$ such that
$C(x_i-x_j)\propto\mid i-j \mid ^{-\gamma}$ where
$\gamma=1-\beta$. Random sequences with weaker positive
correlation is generated with $0.5<\alpha<1$ which is referred to
fractional Gaussian noise~\cite{fractal_book}. In this case,
$\beta=2H-1=2\alpha-1$ and $\alpha=0.5$ corresponds to
uncorrelated disorder or white-noise.

It is commonly usual to apply Fourier filtering
method~\cite{stanley} to generate a sequence with the long-range
correlation. The method is based on a transformation of the
Fourier components {($\theta_{k}$)} of a random number sequence
$\{\theta_{i}\}$. Uncorrelated random numbers $\{\theta_{i}\}$
have a Gaussian distribution. Finally, the inverse Fourier
transformation of the sequence $\{\varepsilon_{k}\}
(=k^{-\frac{\beta}{2}}\theta_{k})$ results to the interested
sequence $\{\varepsilon_{i}\}$. Three landscapes of random data
generated by the mentioned method are shown in
Fig.~\ref{correlated-data} for different correlation exponents.
Clearly, correlation between random potentials leads to a
reduction in the fluctuations of the random distribution. We
should notice that we have checked the invariance of our results
in comparison with those results obtained from random sequences
produced by the mid-point method~\cite{fractal_book}.

Having such configuration of the potentials gives rise a
superlattice consists of two types of graphene as electron-doped
or hole-doped. The doping type of graphene in wells is n-type for
quasiparticles with incident energy $E=hv_{\rm F}/\lambda > 0$
where $\lambda$ is electron wavelength in wells. The doping type
in barriers depends on the potential height $V_i$. In other
words, the $i^{\rm th}$ barrier is hole-doped graphene if
$E<V_i$, while it is electron-doped if $E>V_i$. Accordingly, the
type of doping in barriers is also altered randomly.
\bfig
\includegraphics[width=9 cm]{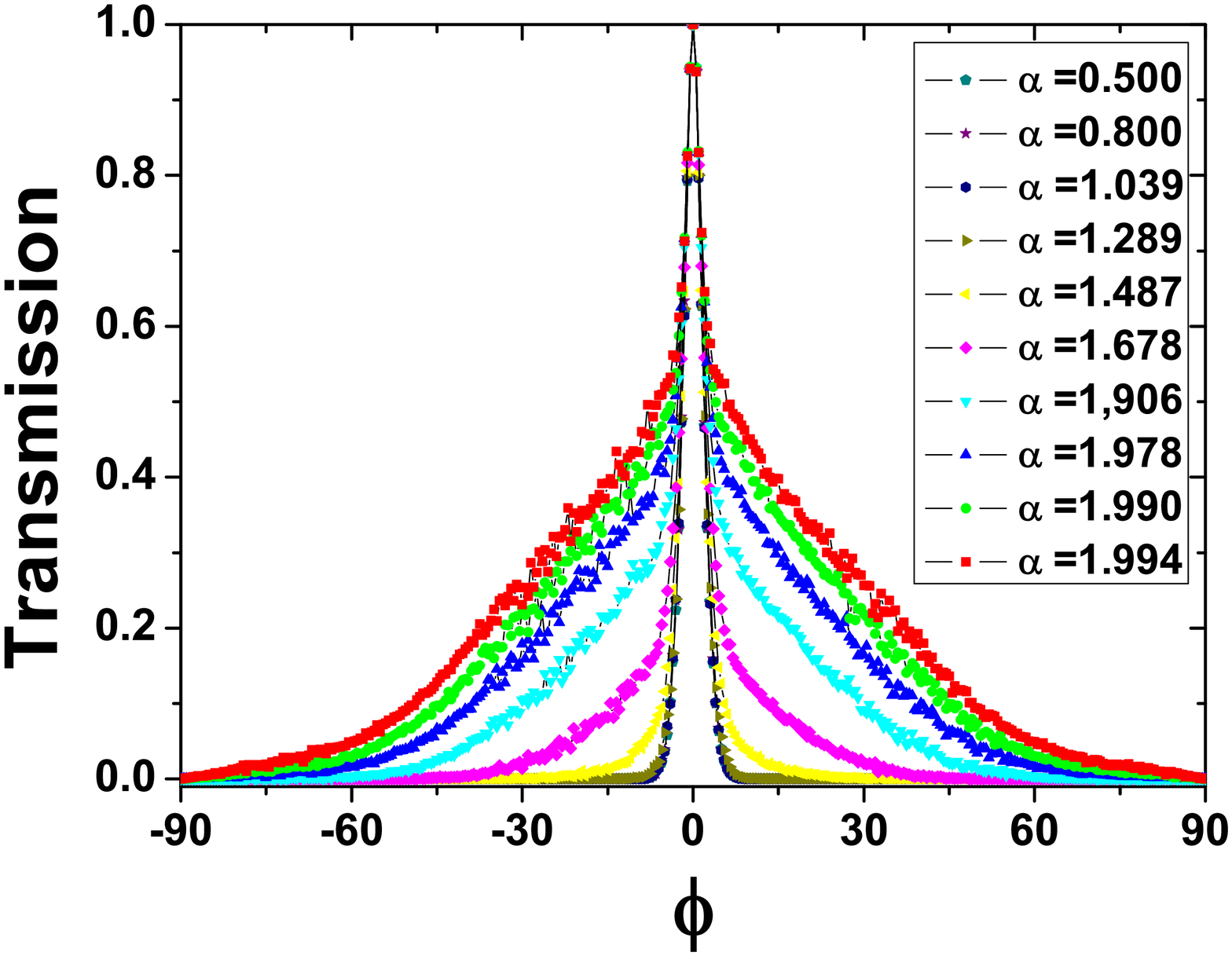}
\caption{Transmission in terms of the incident angle hitting to
graphene superlattice with various correlation exponents
$\alpha$. The number of the barriers is $N=1000$. Averaged potential
of the barriers and energy of the incident electrons are $\langle
V\rangle=200 {\rm meV}$ and $E=50{\rm meV}$, respectively. Here,
since $\xi=-3$, and transmission is calculated in the presence of
purely propagating modes.}\label{transmission-angle} \efig
We assume that the angle of electrons incidence to superlattice
is $\phi=\varphi_1$ along the $x$ axes. However, the transfer
matrix takes the angle of each region into account separately. The
general solution of Eq.~(\ref{dirac}) results in the following
spinor for the $i^{\rm th}$ region;

\begin{eqnarray}\psi(x,y)= a_i \left\{ \begin{array}{c} 1
\\ \\ s_i e^{i\varphi_i}\end{array} \right\}e^{i(k_{ix}x+k_yy)}\nonumber
\\ +b_i \left\{ \begin{array}{c} 1
\\ \\ s_i e^{i(\pi-\varphi_i)}\end{array} \right\}e^{i(-k_{ix}x+k_yy)} \label{spinor}
 \end{eqnarray}
where $a_i$ and $b_i$ are the transmission and reflection amplitudes,
respectively. Other parameters in the spinor are

\begin{eqnarray} s_i={\rm sgn}(E-V_i), \,\,\,\,\ k_y=k_{\rm F}^i\sin(\varphi_i)=k_{\rm F}^i\sin\phi \nonumber
\\  k_{ix}=\sqrt{((E-V(x))/\hbar v_{\rm F})^2-k_{y}^2}, \,\,\,\,\,\
\varphi_i=\arctan(k_y/k_{ix}). \label{parameters}
 \end{eqnarray}

If the energy of the incident electrons being close to $V_i$
value in the $i^{\rm th}$ barrier, $k_{ix}$ will become imaginary
value in some angles, resulting in an evanescent mode. In
disordered graphene superlattice, evanescent modes emerge when
$E\simeq \langle V \rangle$. The transfer matrix is extracted by
the continuity of the wave functions at the junction interfaces.
It can make a relation between wave functions of two sides of a
step potential from the $i^{\rm th}$ region to the $(i+1)^{\rm
th}$ like \beq
\begin{pmatrix}
a_{i+1}\\
b_{i+1}
\end{pmatrix}
= M_{i+1,i}
\begin{pmatrix}
a_{i}\\
b_{i}
\end{pmatrix}
\eeq
where the transfer matrix $M_{i+1,i}$ is;

\beq
M_{i+1,i}=\begin{pmatrix}
m_{11} & m_{12} \\
m_{12}^{*} & m_{11}^{*}
\end{pmatrix}\eeq
and the matrix elements of $M$ are as follows;

\begin{eqnarray}
m_{11}&=e^{i(k_{ix}-k_{(i+1)x})x_i}(\frac{s_{i+1}e^{-i
\varphi_{i+1}}+s_i e^{i\varphi_i}}{2s_{i+1}\cos\varphi_{i+1}})&
\\ \nonumber
m_{12}&=e^{-i(k_{ix}+k_{(i+1)x})x_i}(\frac{s_{i+1}e^{-i
\varphi_{i+1}}-s_i e^{-i\varphi_i}}{2s_{i+1}\cos\varphi_{i+1}})&
\\ \nonumber
\end{eqnarray}

The current in the $x-$direction and in the $i^{\rm th}$ region can
be derived as the following;

\beq J_x^i=v_{\rm F}\psi^{\dag} \sigma_x \psi = 2v_{\rm F} s_i \cos \varphi_i
(\mid a_i \mid^2-\mid b_i \mid^2)\eeq

Current conservation between regions $i^{\rm th}$ and $j^{\rm
th}$ implies on

\beq \mid a_{j} \mid^2-\mid b_{j} \mid^2=Det[M_{j,i}](\mid a_i
\mid^2-\mid b_i \mid^2) \eeq
where
 \begin{eqnarray}
M_{j,i}=M_{j,j-1}M_{j-1,j-2}...M_{i+1,i} \\
\nonumber Det[M_{j,i}]=s_i\cos\varphi_i/s_{j}\cos\varphi_{j}
\label{product}\end{eqnarray}

The total transfer matrix which makes a relation between incident and
transmitted wave functions is a series product of the transfer matrices
arising from each interface. For $N$ regions incorporating of the barriers
and wells, regarding to Eq.(\ref{product}), the matrix is
defined as $P=M_{N,1}$.
\bfig
\includegraphics[width=9 cm]{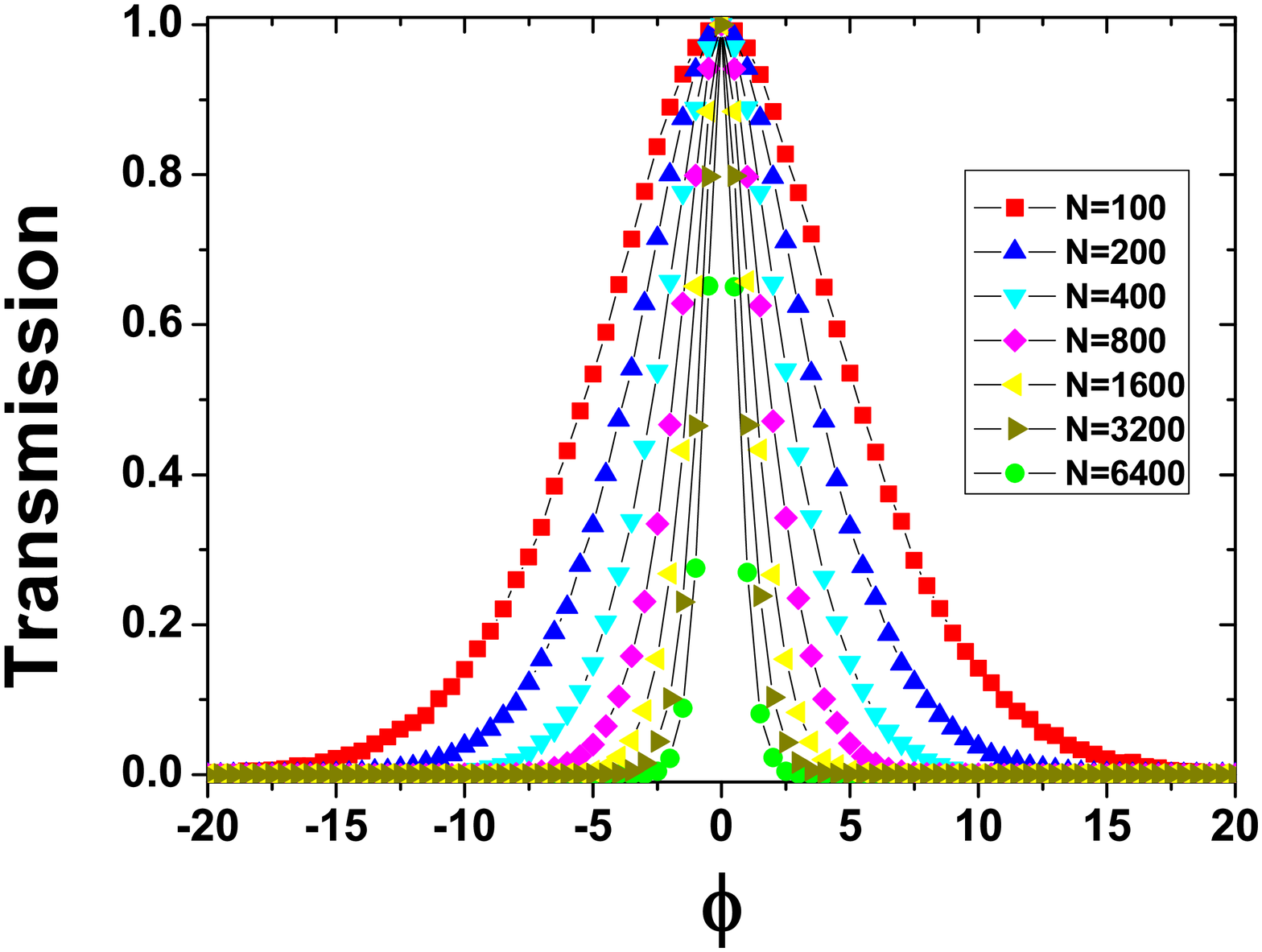}
\caption{Transmission as a function of the incident angle through graphene
superlattice with long-range correlated random potentials for various barrier numbers $N$. Here $\alpha=1.102$ and
$\sigma=0.1$. }\label{transmission-number} \efig

If the first and last regions of superlattice are electron-doped
graphene, transmission probability for $(N-1)/2$ barriers can be
calculated by means of the product matrices for $b_N=0$ as the
following

\beq
T(E,\phi)=\frac{J_{out}^N}{J_{in}^1}=\frac{1}{Det[P]}\left|\frac{a_N}{a_1}\right|^2~,
\eeq
where $J_{out}^N$ and $J_{in}^1$ are out- and inflowing currents, respectively. Because the
configuration of the potential barriers is considered such a way that $s_1=s_N$
and $\varphi_1=\varphi_N=\phi$, the conservation of the current
between the first and last regions implies that $Det[P]=1$. Therefore,
the transmission formula can be simplified as $T(E,\phi)=1/ \mid
P_{22}\mid^2$ where $a_N/a_1=Det[P]/P_{22}$. Finally, using
Landauer-B\"{u}ttiker formula~\cite{datta} and an angularly averaging, the conductance is
obtained by the following integration.

\beq G=G_0\int_{-\pi/2}^{\pi/2}T(E,\phi)\cos(\phi)d\phi
\label{conductance}\eeq where $G_0=e^2mv_{\rm F}W/\hbar^2$. Here $W$ is
the finite width of graphene ribbon along the $y-$direction.
\bfig
\includegraphics[width=9 cm]{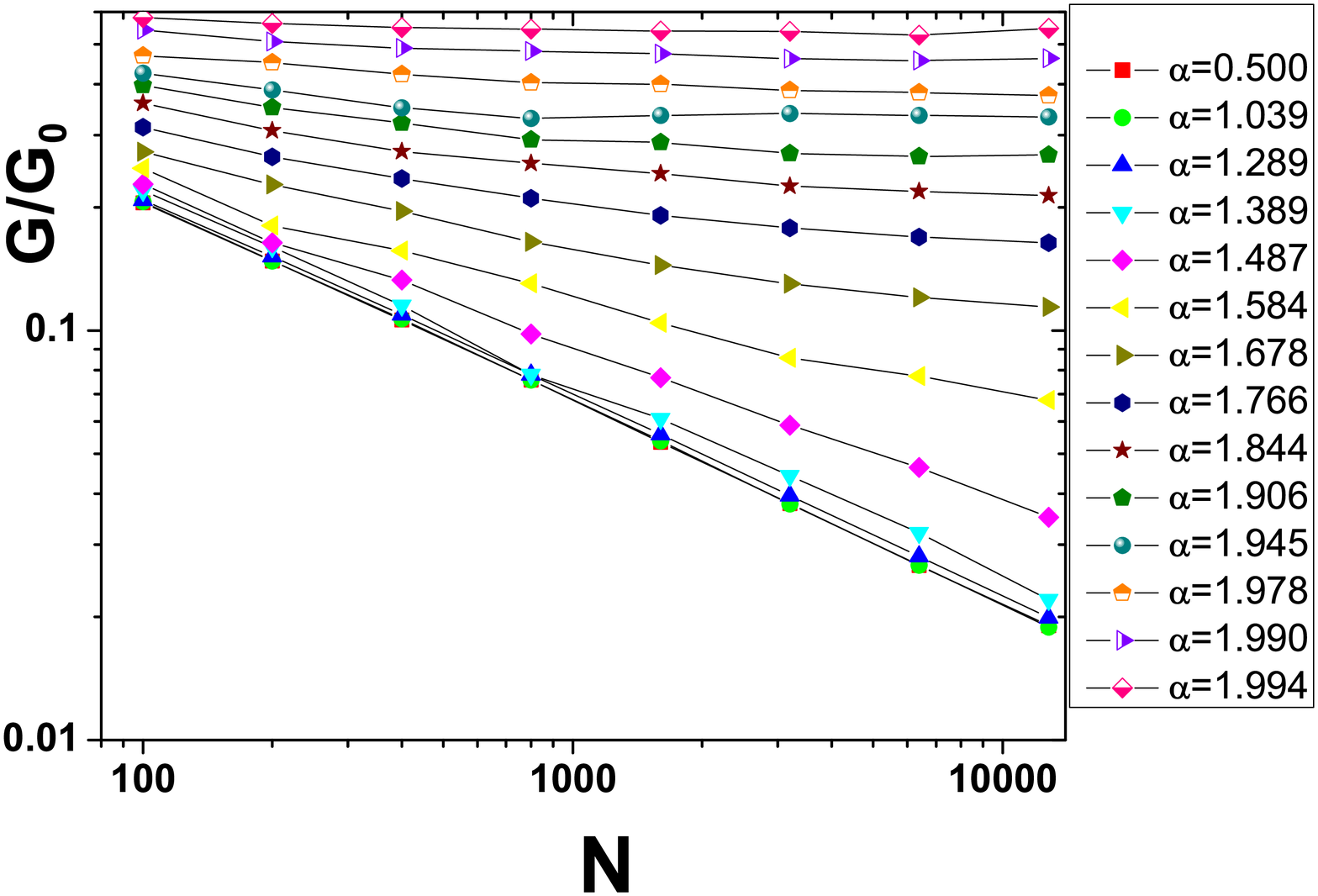}
\caption{Conductance through graphene superlattice with long-range
correlated random potentials as a function of barrier number $N$
for $\sigma=0.1$. This is a finite size scaling for different
correlation exponents $\alpha$. Here, $\langle V\rangle=200 {\rm
meV}$ and $E=50{\rm meV}$.}\label{conductance-number} \efig

\section{Results and discussion}
\subsection{Phase transition}

Let us first calculate the transmission probability and study the
electronic properties of disordered graphene superlattices. The
transmission of electrons hitting to a disordered graphene
superlattice as a function of the incident angle is shown in
Fig.~\ref{transmission-angle} for several values of correlation
strengths characterizing with the correlation exponent $\alpha$.
In all calculations, barrier and well widths are considered to be
$D=50{\rm nm}$ and $L=30 {\rm nm}$, respectively. Moreover, we
assumed that the energy of the charge carriers and the averaged
potential of barriers being $E=50{\rm meV}$ and $\langle
V\rangle=200 {\rm meV}$, respectively. The wavelength of the
incident electrons is thus $\lambda\cong 83{\rm nm}$. Therefore,
with such parameters we surely conclude that the transmission of
the charge carriers shown in Fig.~\ref{transmission-angle} (with
$\xi=(E-\langle V\rangle)/E=-3$) is a purely propagating mode. As
shown in this figure, the transmission of the electrons hitting
to superlattice with large angles increases by increasing
correlation between the random potentials of the barriers. In
other words, applying correlation between random potentials of
the barriers causes to extend the angular window of the
conducting mode around the normal incidence. This effect is in
contrast with those results obtained in uncorrelated potentials
of the barriers in which the transmission of the massless carriers
is suppressed for the large incident angles except at $\phi=0$.
Perfect transmission at normal incidence can be described by the
Klein tunneling. Similarly, in the correlated case, by increasing
disorder strength and also the number of the barriers, the
transmission of the quasiparticles is suppressed at all ranges of
the incident angle except at $\phi=0$.
Fig.~\ref{transmission-number} shows the suppression of the
transmission at large incident angles when the number of the
potential barriers increases.

\bfig
\includegraphics[width=9 cm]{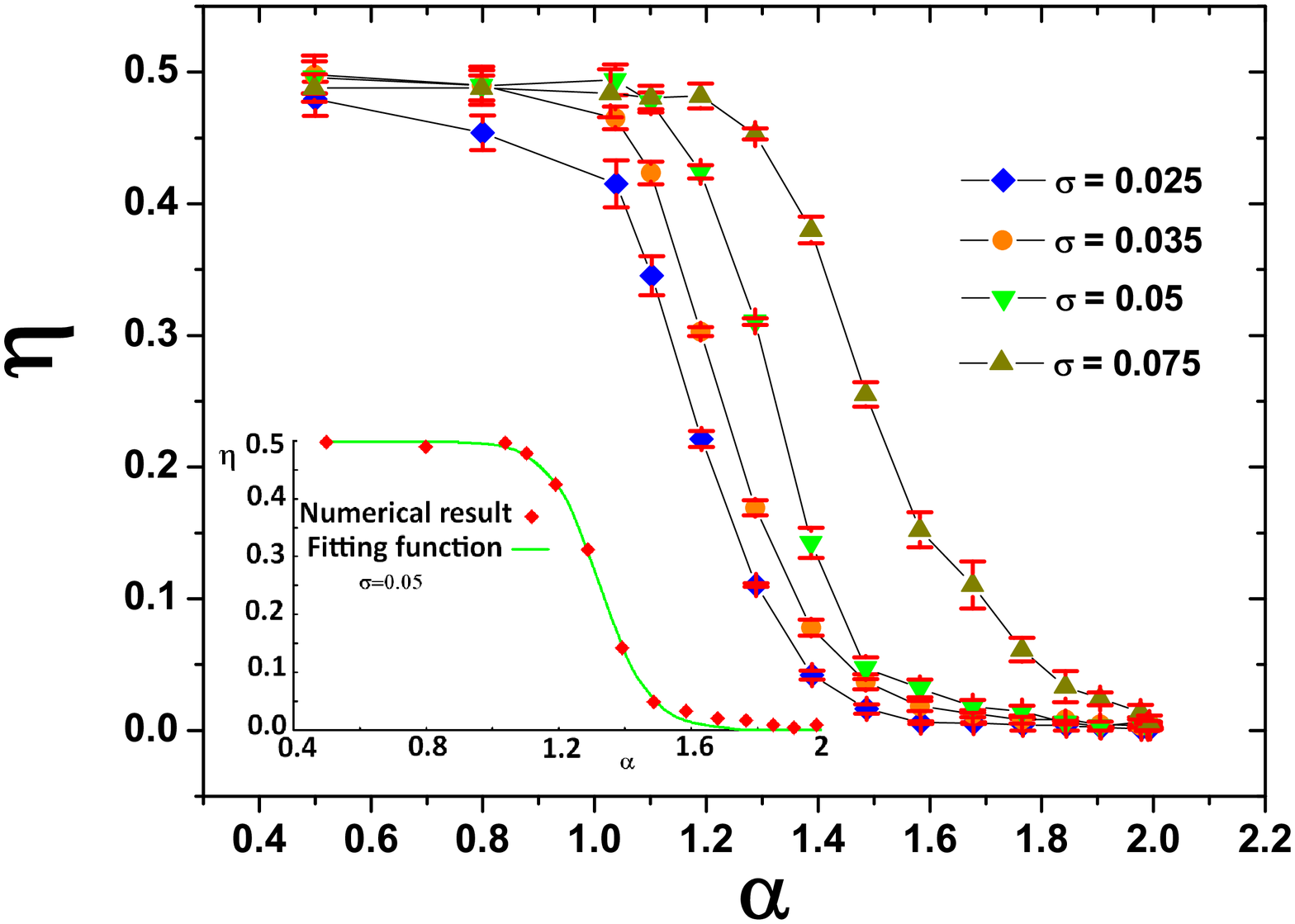}
\caption{Exponent of the conductance in a power law form
$G/G_0\propto N^{-\eta}$ as a function of correlation exponent
$\alpha$. In the inset the fitting of a Fermi-Dirac like
function (Eq.\ref{Fermi-Dirac}) to numerical
data is shown.}\label{etha-hurst} \efig

In linear regime, the conductance is proportional to angularly
averaged transmission projected along the current direction. To
understand how the correlation between random potentials affects
on the transport properties of graphene superlattice, we
calculated the size dependence of the conductance for the various
values of the correlation exponents and results are plotted in
Fig.~\ref{conductance-number}. It is clear from the figure that
there is a  critical correlation exponent value of $\alpha$ such
that for $\alpha<\alpha_{cr}$, the conductance decreases with
increasing system size, while it goes to a constant value for
$\alpha>\alpha_{cr}$. In other words, a consequence of applying
long-range correlation is the emergence of a phase transition from
the insulating to metallic states. It is worthwhile noting that
this phase transition is not a finite size effect. The conductance
decreases with the number of the barriers as a power law
behavior. The following function is fitted to a log-log plot of
the conductance.

\beq \frac{G}{G_0} \propto N^{-\eta(\alpha,\sigma)} \eeq

where $\eta$ as functions of $\alpha$ and $\sigma$ decreases by
increasing the correlation exponent. Furthermore, disorder
strength strongly suppresses the conductance such that the
critical correlation exponent which implies on the emergence of a
phase transition, increases with the disorder strength.
\bfig
\includegraphics[width=9 cm]{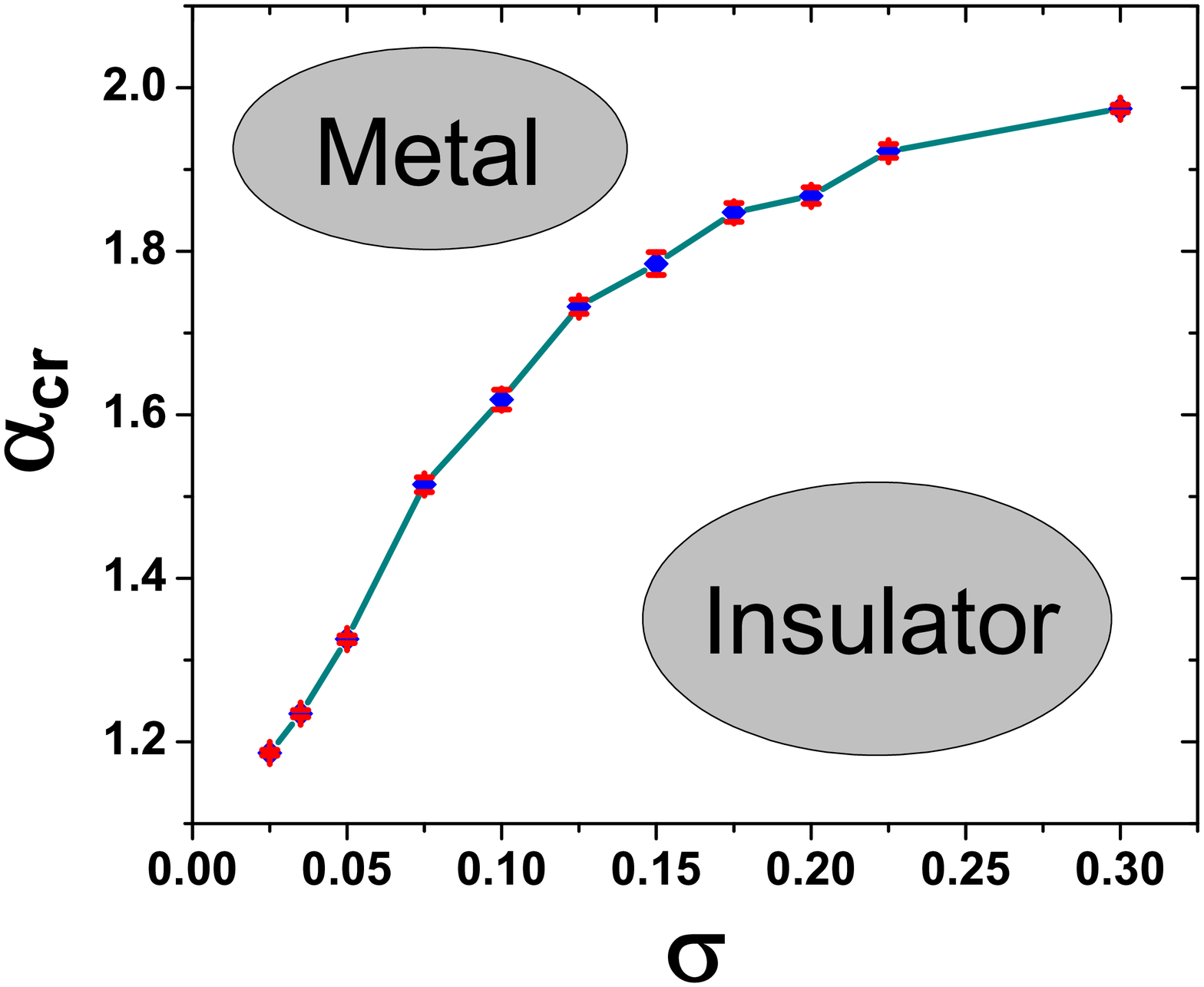}
\caption{Metal-to-insulator phase diagram. The
critical correlation exponent, $\alpha_{cr}$ increases with
disorder strength, $\sigma$.}\label{sigma-hcr} \efig

Conductance through disordered potentials of barriers having
correlation exponent $\alpha=0.5$ decays with the barriers number
as $N^{-0.5}$. The same decaying of the conductance was reported in
Ref.~[\onlinecite{asgari}] for graphene superlattice with white
noise potentials distributed on the barriers. However, applying a
long-range correlation between random potentials facilitates
the conductance through graphene superlattice.

A power-law fitting for the conductance is shown in
Fig.~\ref{conductance-number}. It is determined a dependency of
$\eta$ on the correlation exponent $\alpha$ and disorder strength
$\sigma$ which is represented in Fig.~\ref{etha-hurst}. For the
sake of having the critical correlation exponent function, we have used a fit function
of the exponent function $\eta(\alpha,\sigma)$ in a transition
function such as a Fermi-Dirac function:

\beq
\eta(\alpha,\sigma)=\frac{\gamma}{e^{\beta(\alpha-\alpha_{cr}(\sigma))}+1}
\rightarrow\left\{ \begin{array}{c} 0
\,\,\,\,\,\,\,\,\,\,\,\,\,\,\,\,\,\ \alpha >> \alpha_{cr}
\\ \\ \gamma  \,\,\,\,\,\,\,\,\,\,\ \alpha << \alpha_{cr}\end{array} \right.
\label{Fermi-Dirac}
 \eeq
where $\beta$ and $\alpha_{cr}$ are fitted parameters and
$\gamma=0.5$ for $\xi=-3$. The range that $\eta(\alpha)$
decreases from the value of $\gamma$ to zero is related with the
inverse of $\beta$. The emergence of a transition from insulating
to metallic phase corresponds to the variation of $\eta$ from
$\gamma$ to $0$. Now, we provide a phase transition diagram in
which the critical correlation exponent depends on disorder
strength. Fig.~\ref{sigma-hcr} shows that the critical correlation
exponent increases when disorder strength increases up to $\sigma=0.3$. Roughly speaking,
the system remains in a metallic phase when the correlation is very long range
at low disorder strength whereas it turns out to be an insulator at
large disorder strength values and at mid-range correlation.

\bfig
\includegraphics[width=8 cm]{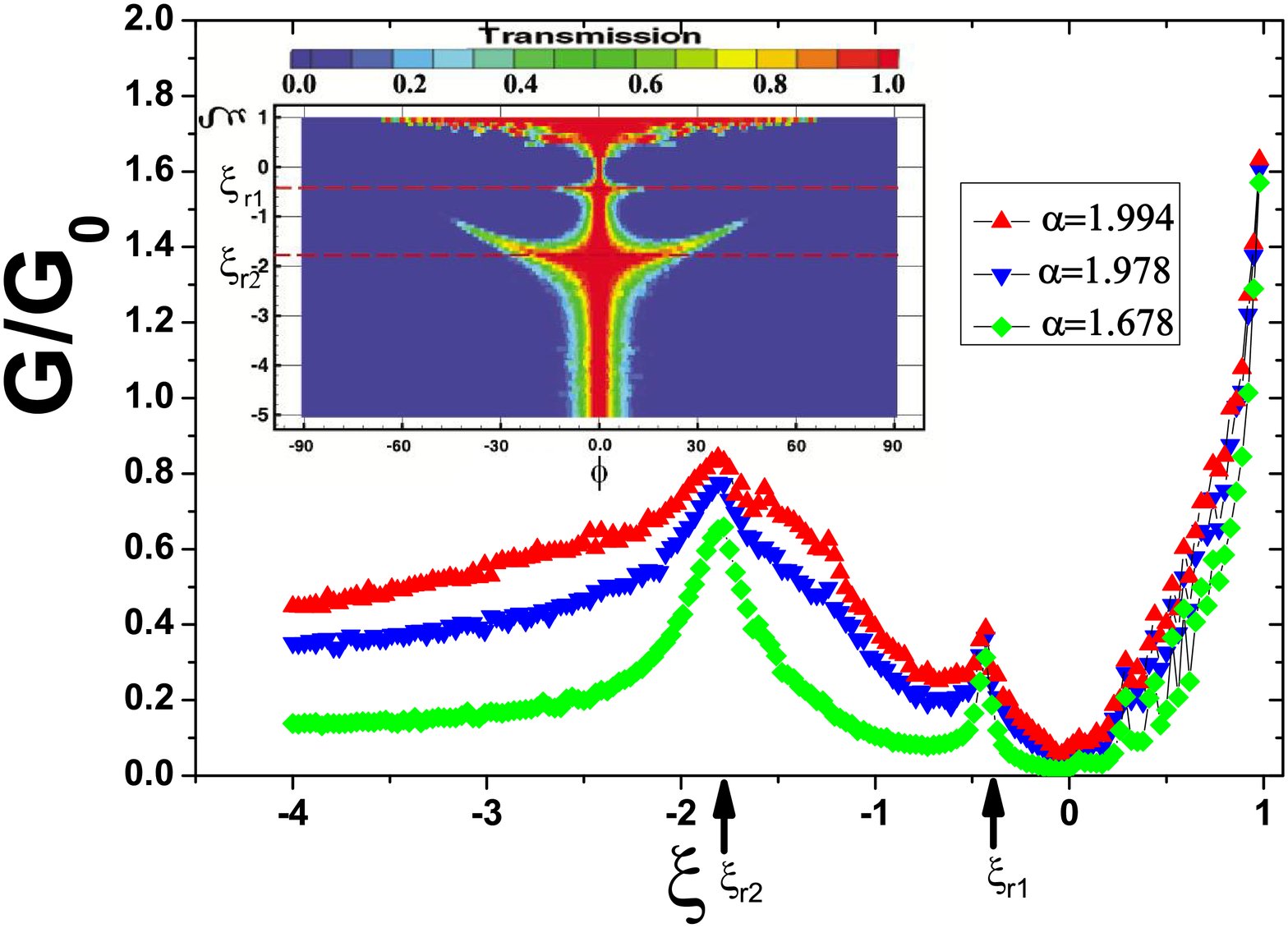}
\caption{Functional of the conductance in terms of $\xi$ which
represents Fermi energy. There are two peaks in conductance correspond to resonant
states. In the inset 3D contour plot of the transmission in the plane of
$\xi$ and the incident angle. Resonant states corresponds to more
angularly open a domain for transport. Here $N=800$ and $\sigma=0.1$.}\label{conductance-xi} \efig

\subsection{Energy range of the phase transition}

To provide a fascinating experimental manifestation of the phase
transition, it is significant to demonstrate that the phase
transition can exist in a continuum range of energies not just at
some discrete energies.

Let us now concentrate on the conductance behavior in the
different range of the Fermi energy. For a single barrier on
graphene, in the range of $-1<\xi<1$, it is proved that both the
evanescent and propagating modes coexist~\cite{proximity}. Out of
this range, all states are fully in the propagating modes. By
considering this fact, we investigate the conductance as a
function of $\xi$ through a disordered graphene superlattice for
several correlation exponent values and the results are shown in
Fig.~\ref{conductance-xi}. It can be seen that the same behavior
as a single barrier case is appeared in the different ranges of
$\xi$. Contribution of the evanescent modes in the conductance is
dominant in $\xi=0$ and therefore the conductance is suppressed
in this point. For $|\xi|\rightarrow 1$, the contribution of the
propagating modes becomes dominant and thus the conductance
increases. For $\xi <0$, the conductance oscillates at resonant
states which originates from the perfect tunneling of the charge
carriers near to a normal incidence. A 3D contour plot of the
transmission in terms of $\xi$ and the incident angle (see the
inset Fig.~\ref{conductance-xi}) shows that at resonant states, a
conducting domain of the angles is opened around the normal
incidence. The resonant condition for a single barrier in
graphene is given by $k_{\rm F}D\sqrt{\xi^2-u^2}=n\pi$ where
$u=\sin\varphi$. Distance of the resonant states can be extracted
from expanding the resonant condition for normal incidence
$u\ll\xi$. In this case, two sequential resonant states have a
distance like $\Delta\xi_r\cong\frac{\pi}{k_{\rm
F}D}(1-\frac{u^2}{2\xi^2})$. As a result, the period of the
conductance reduces with $D$, $L$ while enhances with $|\xi|$. In
the limit of $\xi\rightarrow1$ which it means $E\gg~<V>$, the
conducting channels are opened for the whole range of the angles,
$T(\phi)=1$, and thus by using Eq.~(\ref{conductance}), we then
get $\lim_{\xi\rightarrow1}G/G_0=2$.
\bfig
\includegraphics[width=8 cm]{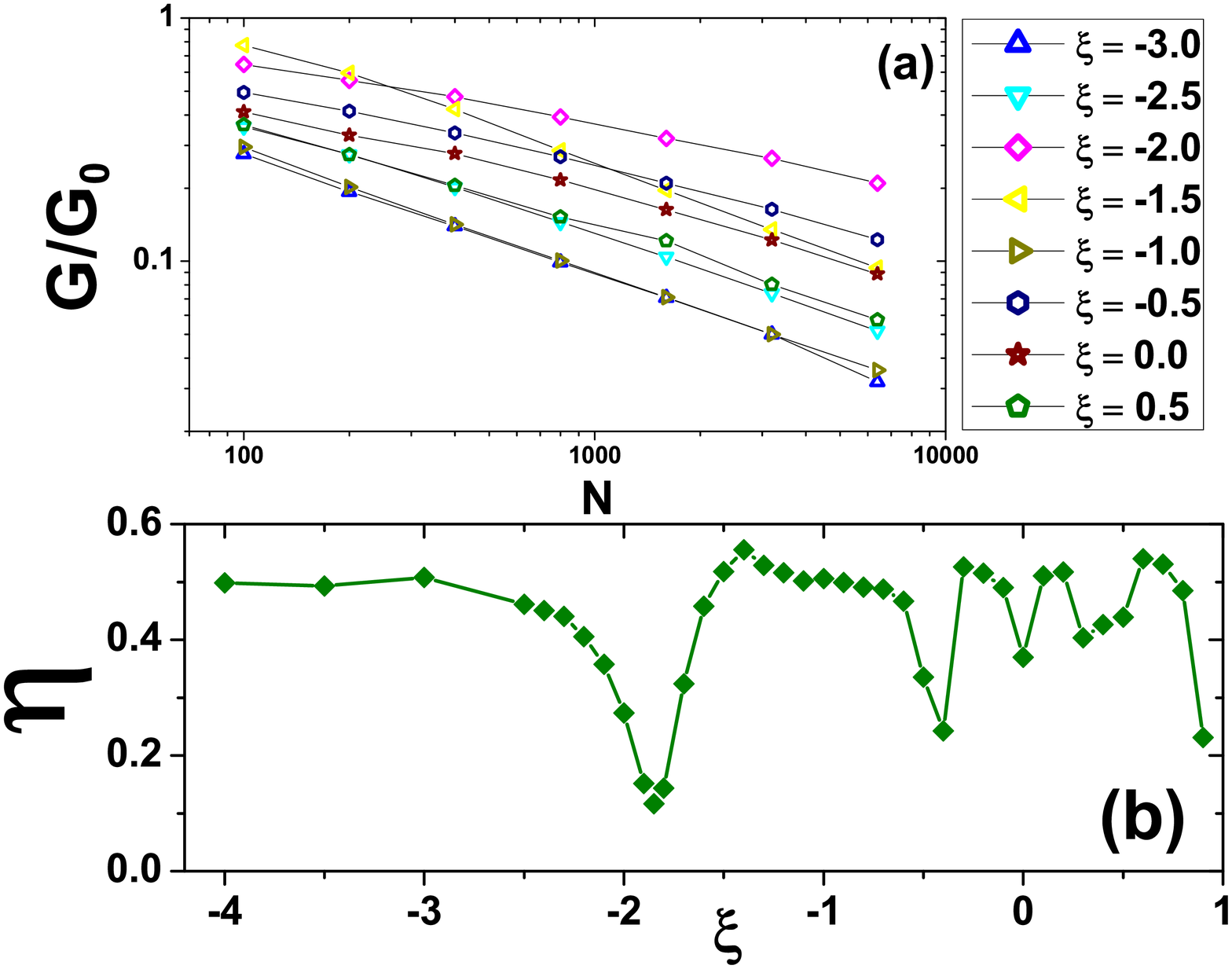}
\caption{a) Conductance in terms of the barrier numbers for graphene
superlattice with white-noise disorder on potentials of the barriers
and for different values of the Fermi level $\xi$. b) Exponent of
the conductance ($\eta$) in a power-law form $G/G_0\propto N^{-\eta}$
as a function of $\xi$.}\label{conductance-N-xi} \efig

By applying a long-range correlation between the potentials of
the barriers, Fig.~\ref{conductance-xi} indicates that the conductance
shows an enhancement in some ranges of the energy compared to
the resonant states and the range including the evanescent modes
$\mid\xi\mid<1$. For clarifying that, firstly we
investigate the conductance suppression with the number of the barriers
over all ranges of the Fermi energy for a sequence of white noise
disorder distributed on the potentials of the barriers. Results are
represented in Fig.~\ref{conductance-N-xi}a which show a power-law
form of the conductance in terms of the barrier numbers for different
values of $\xi$. It is significant to understand how the exponent of $\eta$
varies with the Fermi energy. The slope of the lines in log-log plot of
Fig.~\ref{conductance-N-xi}a for different $\xi$ are indicated
in Fig.~\ref{conductance-N-xi}b. It is clear that close to
resonant states $\xi_r=-1.85,-0.45$, the exponent of $\eta$
reduces from the value of 0.5 to its resonant values
$\eta_r=0.12,0.25$, respectively. As a consequence, close to
resonant states, the suppression of the conductance which is induced by
random potentials of the barriers is much weaker than other states.
The resonant value of $\eta_r$ decreases when $\xi$ goes away from the
region including the evanescent modes $\mid\xi\mid<1$.

Our numerical calculations demonstrate that the insulator-to-metal
phase transition occurs in all ranges of the energy. It is clear from
Fig.~\ref{eta-alpha-xi} that the phase transition appears not
only at the propagating and resonant states, but also in the fully
evanescent mode $\xi=0$. In Fig.~\ref{eta-alpha-xi},
$\eta(\alpha=0.5)$ decreases to $0.3$ at $\xi=-2$ which is close
to the second resonant state shown in Fig.~\ref{conductance-xi}.
Another exception around the resonant states is the width of
function $\eta(\alpha)$ which increases at the resonant states.
Therefore, at the resonant states, the transition from insulating to
metallic phase is smooth along the correlation exponent. The inset
Fig.~\ref{eta-alpha-xi} shows small fluctuations of the critical
correlation exponent as a function of $\xi$. Accordingly, the phase
transition is a general behavior of all ranges of the energy.

\bfig
\includegraphics[width=9 cm]{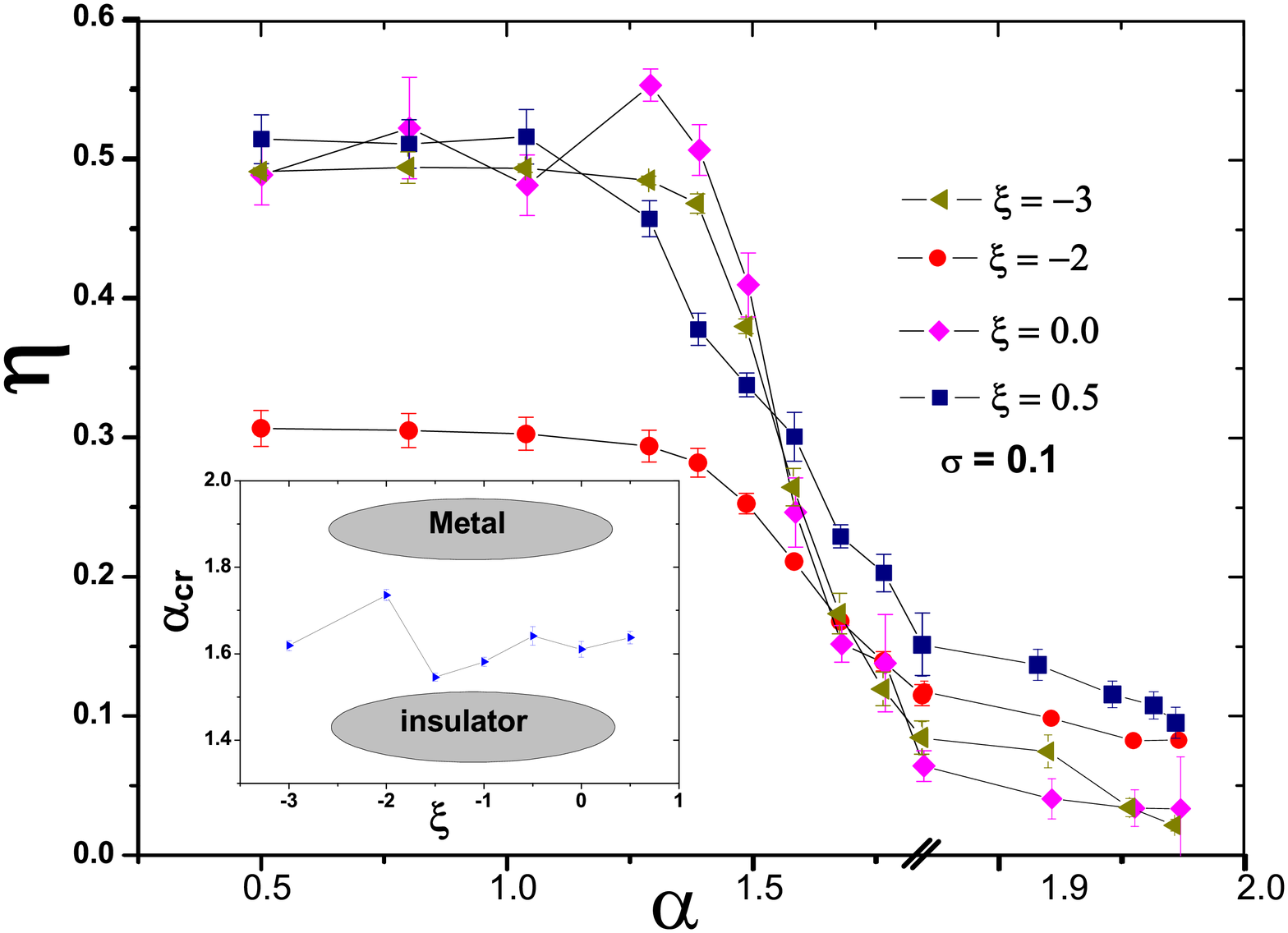}
\caption{Exponent of the conductance in a power law form
$G/G_0\propto N^{-\eta}$ as a function of the correlation exponent
$\alpha$ for different values of $\xi$. In the inset the
metal-to-insulator phase transition in all ranges of
the energy.}\label{eta-alpha-xi} \efig

\subsection{Resonance in the conductance}
Now, we study the effect of the long-range correlated disorder on
the resonance phenomena seen in the conductance. Resonance in the
conductance of graphene superlattice with white noise disorder
distributed on width of the barriers has been studied
before~\cite{asgari}. Fig.~\ref{conductance-L-D} shows the
conductance oscillations as a function of the barrier width ($D$)
and also distance between barriers ($L$) for several values of
the correlation exponent. Apparently, the application of the long
range correlation between potentials of the barriers increases
the conductance for all ranges of $D$ and $L$. Moreover, as shown
in Fig.~\ref{conductance-L-D}a, the conductance tends to a
constant value after some oscillations in thin barriers
independent of the barrier width. In fact, the transmission of
the quasiparticles hitting to graphene superlattice at large
incident angles is suppressed for the wide barriers, and thus
only transmission arising from the Klein tunneling around the
normal incident contributes into the conductance integration
given by Eq.~(\ref{conductance}). The same behavior occurs in the
case of $N=1$. Configuration average of the transmission through
one barrier~\cite{RMP} depends on the width by functions of
$<\sin^2(k_xD)>_{C.A.}$ and $<\cos^2(k_xD)>_{C.A.}$ where $k_x$
is a random parameter. It is trivial that the average of the
transmission and consequently the conductance become independent
of the width when $D$ value increases. On the other hand, by
applying of the randomness on $k_x$, the resonant condition
$k_xD=n\pi$ in wide barriers can not be satisfied, and there is
thus no longer the resonant peak in the conductance.
\bfig
\includegraphics[width=9 cm]{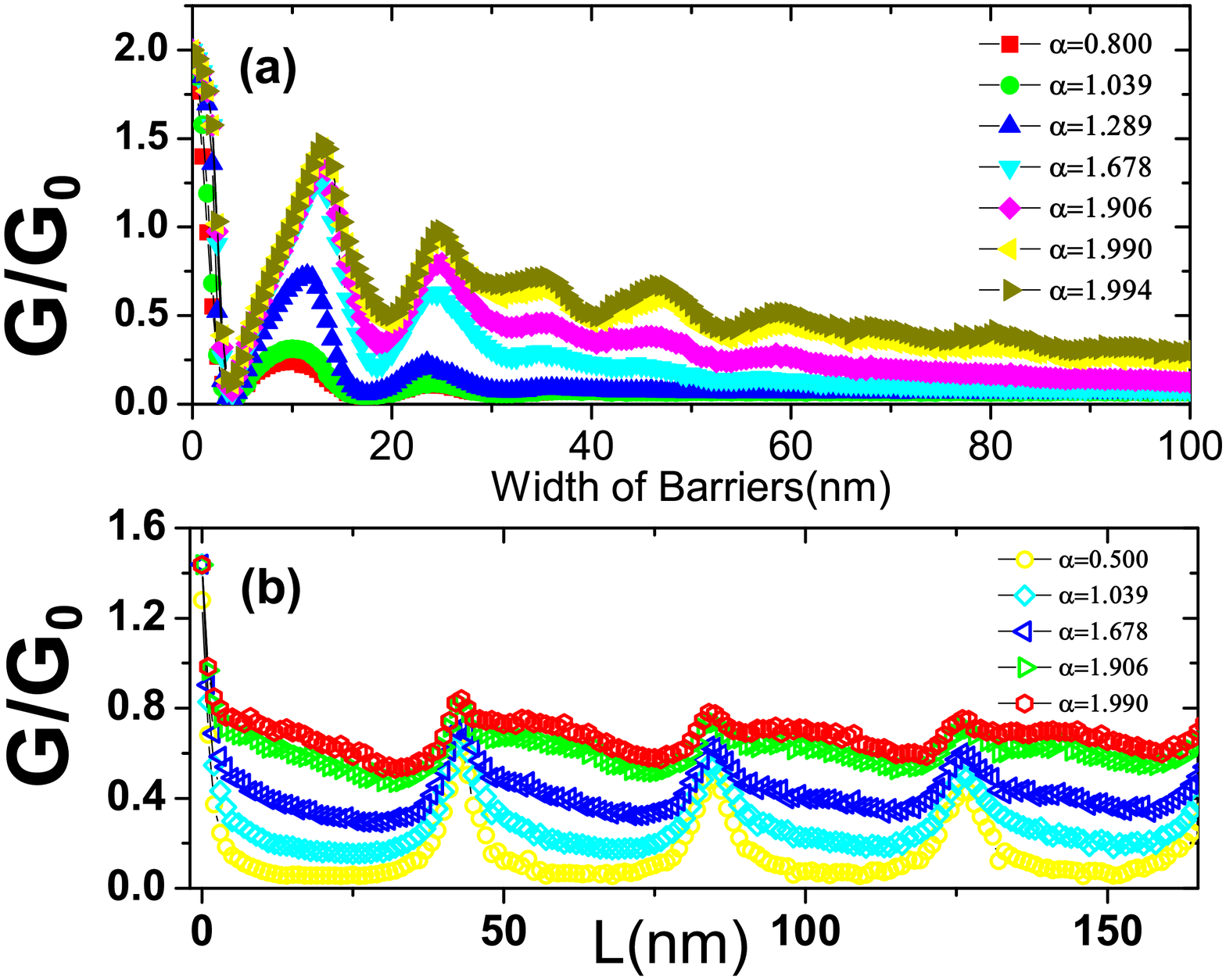}
\caption{Conductance oscillations in terms of a) the width ($D$)
of the barriers b) the distance ($L$) between barriers for different
correlation exponents. Here
$N=800$ and $\sigma=0.1$.}\label{conductance-L-D} \efig

Fig.~\ref{conductance-L-D}b shows the conductance oscillations as a
function of $L$. In this case, the decaying of the oscillations is much
weaker than the conductance oscillations with the barrier width. In
fact, since there is no disorder in the wells, in a fixed barrier
width, the resonant condition affects much less than the barrier resonant
condition.

\section{Conclusion}

By using the transfer matrix method, we investigate the
conductance through a graphene superlattice with long-range
correlated disorder distributed on potentials of the barriers.
Applying a correlation between potentials opens angularly domain
window of the conducting channels in competition with the factor
of the disorder strength which suppress the transmission at large
incident angles. As a result, the conductance increases with the
correlation between the potentials of the barriers gives rise a
metallic phase. We obtain a phase transition diagram in which the
critical correlation exponent for such a phase transition depends
strongly on disorder strength and slightly on the energy of the
incident particles. At resonant states, the suppression of the
conductance with the number of the barriers is much less than
other states. Our finding for the dc conductance of the graphene
superlattices should be important to the design of electronic
nanodevices based on graphene superlattices.

\end{document}